\documentclass[preprint,12pt]{elsarticle}

\usepackage{amssymb}
\usepackage{siunitx}
\usepackage{xcolor}
\usepackage{cancel}
\usepackage{hyperref}

\usepackage{siunitx}
\DeclareSIUnit\angstrom{\text{Å}} 

\journal{Electrochimica Acta}

\begin{document}

\begin{frontmatter}


\title{Machine-Learned Interatomic Potentials for Predicting Physicochemical Properties of Molten Metal–Salt Systems for Calcium Electrolysis.}

\affiliation[Skoltech]{organization={Skolkovo Institute of Science and Technology},
            addressline={Skolkovo Innovation Center, Bolshoy Bulvar, 30}, 
            city={Moscow},
            postcode={143026},
            country={Russia}}
\affiliation[DigitalMaterials]{organization={Digital Materials LLC},
            addressline={Odintsovo, Kutuzovskaya str. 4A}, 
            city={Moscow Region},
            postcode={143001},
            country={Russia}}
\affiliation[IHTE]{organization={Institute of High Temperature Electrochemistry, Ural Branch of the Russian Academy of Sciences},
            addressline={Academicheskaya Street, 20}, 
            city={Yekaterinburg},
            postcode={620066},
            country={Russia}}

\author[Skoltech,DigitalMaterials]{M. Polovinkin}          
\author[Skoltech,DigitalMaterials]{N. Rybin}
\author[DigitalMaterials]{D. Maksimov}
\author[IHTE]{F. Valiev}
\author[IHTE]{A. Khudorozhkova}
\author[IHTE]{M. Laptev}
\author[IHTE]{A. Rudenko}

\author[Skoltech,DigitalMaterials]{A. Shapeev}

\begin{abstract}

The design of efficient electrolysis devices for pure metal production requires accurate data on the properties of the melts used in the process. This work focuses on two key systems for calcium production: the molten Ca-Cu alloy and the CaCl$_2$-KCl electrolyte. High-temperature experiments are often expensive and time-consuming; however, we demonstrate that molecular dynamics (MD) simulations driven by machine-learned Moment Tensor Potentials (MTPs), trained on highly accurate density functional theory data, offer an effective and accurate alternative. Our MTP-driven MD simulations accurately reproduce the structural, thermodynamic, and transport properties across a range of temperatures and compositions relevant to electrolysis systems. We report calculated densities, radial distribution functions, heat capacities, thermal conductivities, ionic conductivities (for the electrolyte), viscosities, and diffusion coefficients, with deviations from experimental data within 20\%. The strong agreement between calculations and experiments validates the proposed approach, establishing a robust framework for the computational exploration and optimization of liquid systems in metallurgical applications.
\end{abstract}

\begin{highlights}
\item Moment Tensor Potentials for the molten Ca-Cu alloy and the CaCl$_2$-KCl molten salt are developed for the first time
\item The Moment Tensor Potential for the Ca-Cu alloy is compositionally transferable  
\item The same potentials are used to calculate structural, thermodynamic, and transport properties of the systems
\item The ionic conductivity of the CaCl$_2$-KCl molten salt predicted by Moment Tensor Potential molecular dynamics agrees with experiment
\end{highlights}


\end{frontmatter}

\section{Introduction}
\label{sec:intro}

Metallic calcium is a critical element with a wide range of industrial applications, including steelmaking, battery production, and the synthesis of advanced alloys and magnetic materials \cite{alloys_Naumova2018,Ca_rev_Hluchan2006,CaCu_exp_Zaikov,CIB_Gummow2018}. As demand grows, improving calcium production methods is becoming increasingly important. The primary industrial route for producing high-purity calcium is molten salt electrolysis. This process involves two high-temperature liquid phases: a molten Ca-Cu alloy acting as a liquid cathode and a CaCl$_2$-based molten salt electrolyte \cite{Ca_Zaikov2022}.

Optimizing the efficiency of production processes is a key industrial objective. Multiphysics modeling using digital twins can greatly assist in these efforts \cite{DT_VERGARA2025}. However, digital twins require precise knowledge of the structural, thermodynamic, and transport properties of both the electrolyte and the liquid cathode under operating conditions. This includes the temperature dependence of key properties such as density, viscosity, diffusion coefficients, thermal conductivity, heat capacity, and ionic conductivity.

To the best of our knowledge, a gap exists in the available data for several physicochemical properties of the liquid Ca-Cu alloy and CaCl$_2$-KCl molten salt. The temperature dependence of density has been measured for several compositions of molten Ca-Cu alloys \cite{calcium_Hiemstra1997,CaCu_exp_Zaikov,copper_Assael2010}. In contrast, structural investigations have been limited to pure Ca and Cu melts \cite{CaRDF_Waseda1974,waseda1980structure}, leaving a gap in our understanding of the alloy structure. The heat capacity of Ca-Cu alloys remains unknown, and reported diffusion coefficients and viscosities require refinement, as they are available from only a single source \cite{CaCu_exp_Zaikov} and contain inconsistencies. While the properties of the CaCl$_2$-KCl molten salts have been studied for other compositions \cite{janz1988thermodynamic,PureSalt_TC_SMIRNOV1987,DeepMD_CaKCl_Bu2021}, data for the specific 80:20 mass\% ratio used in production are lacking.

Acquiring these data through high-temperature experiments is often resource-intensive, expensive, and challenging, making such approaches impractical for the large-scale parameter screening—across composition, temperature, and pressure—that industrial optimization demands. Computational approaches offer a viable alternative to experiments \cite{moltensalt_Porter2022}. While {\it ab initio} molecular dynamics (AIMD) can predict physicochemical properties, its computational cost is prohibitively high for sufficient configuration sampling and for the calculation of transport properties. Molecular dynamics (MD) modeling with machine-learning interatomic potentials (MLIPs), including Moment Tensor Potentials (MTPs), offers a favorable trade-off, providing accurate modeling at a fraction of the computational cost \cite{MLIPs_zuo2020, MLIPs_Jacobs2025}. 

In this article, we develop and benchmark a computational methodology employing MTPs to fill data gaps for these systems. First, we construct diverse training sets from density functional theory (DFT) calculations to fit reliable potentials. We then employ MTP-MD to compute a comprehensive set of physicochemical properties for both systems involved in the electrolytic production of calcium. For the Ca-Cu alloy, we address notable gaps in the literature by reporting the composition dependence of heat capacity and by refining inconsistencies in the available viscosity and diffusion data. For the CaCl$_2$-KCl electrolyte, we provide a complete set of properties, including ionic conductivity. To support our computational results, we conducted a series of experiments measuring the density, viscosity, and ionic conductivity of the molten salt. Combined with literature data, these results enable a proper benchmark. 

The computational framework established here is both general and transferable. It can be reliably applied to calculate the physicochemical properties of other molten salts and liquid alloys, providing a powerful tool for the design of new materials and the optimization of production processes.

\section{Methods}
\label{sec:methods}
\subsection{MTP fitting}

To calculate the physicochemical properties, we conducted the MD simulation in LAMMPS \cite{LAMMPS_Plimpton1995,Lammps_Thompson2022}. The MTPs, implemented in MLIP-2 \cite{mlip2_Novikov2021}, were used as the interaction potentials. In the MTP, the energy of an atomic configuration is represented as a sum of energies of local atomic environments. These energies are expanded through a set of basis functions, with linear coefficients determined by fitting energies, forces, and stresses to reference ones. For a comprehensive theoretical description of the MTPs, we refer the reader to the original publications \cite{MTP_Shapeev2016,mlip2_Novikov2021,mlip3_Podryabinkin2023}.

For the systems of the Ca-Cu melt (for all compositions) and the CaCl$_2$-KCl (80:20 mass\%, 28:10 in molar fractions) electrolyte, we developed potentials containing 127 and 343 parameters, respectively. The potential for Ca-Cu liquid alloy was designed to be compositionally transferable. In the MTPs we used the Chebyshev radial basis and a cutoff distance of 5~\si{\angstrom}. Potentials were trained on atomic configurations sampled from the liquid bulk phase, with the simulation cell size ensuring a minimum length of twice the cutoff distance. The MTPs were trained in a multi-stage procedure developed to ensure both accuracy and stability in MD modeling. 

First, for each system, an initial training set was generated from a short 2~ps AIMD simulation. We sampled 150 evenly spaced atomic configurations from the last 1.5~ps of the trajectory. This set of configurations was used to train an initial MTP of level 10. To enhance the stability of the potential in MD, we employed an active learning procedure \cite{AL_Podryabinkin2017}. It consists of performing 100~ps isothermal-isobaric ensemble (NPT) MD modeling, during which atomic configurations with the highest extrapolation grade are automatically added to the training set. 

The initial AIMD produced atomic configurations that were structurally similar. To construct diverse and representative training sets, we performed a "resampling" step using a stable potential. For the Ca-Cu alloy, to ensure transferability across compositions, six separate 60~ps trajectories were run, with the calcium molar fraction ranging from 0.0 to 1.0 in steps of 0.2. For the molten salt, a single 300~ps trajectory was used. From these simulations, 360 (Ca-Cu) and 300 (CaCl$_2$-KCl) atomic configurations were sampled. This approach is effective because the structural relaxation times in these molten systems are short, on the order of 5--10~ps.

Based on the collected diverse training sets, we trained new potentials of level 10. Then, we conducted active learning over a broad temperature range (900–1600~K) to enrich the training set with atomic configurations at various temperatures. Finally, higher-level potentials were fitted to the enriched training set and underwent a final round of active learning to guarantee their stability. This pipeline ensures that the potentials are trained on sets covering a broad chemical space and stable enough to run long MD simulations for the calculation of physicochemical properties.

Reference energies, forces, and stresses for MTP fitting were computed within the DFT framework in the Vienna {\it Ab initio} Simulation Package (VASP) \cite{Vasp_Kresse1999} employing the projector-augmented-wave and the GGA approximation with the PBE functional \cite{PBE_GGA_Perdew1996}. The plane-wave basis kinetic energy cutoff was set to 520~eV and 500~eV for the liquid alloy and molten salt, respectively. A $1\times1\times1$ $\Gamma$-centered k-point mesh was used to sample the Brillouin zone. The width of smearing in the Gaussian method was chosen to be 0.05~eV. The vdW-dispersion correction term was accounted for by the DFT-D3 method of Grimme \cite{D3_Grimme2010,D3_Grimme2011} for the Ca-Cu liquid alloy and by the dDsC method \cite{dDsC_steinmann2011generalized,
dDsC_steinmann2011} for the molten salt. The choice of dDsC was based on the results of our in-house tests conducted on the AlF$_3$-NaF molten salt. Details are provided in Section 1 of the Supplementary Information.

To quantify the accuracy of the MTP in reproducing the DFT results, we computed the root-mean-square errors (RMSE) for energies and forces between the MTP and reference DFT calculations on an independent validation set collected from MD with the final potentials. 

\subsection{Physicochemical properties calculation}

In MTP-driven MD simulations, we used a 1~fs time step and periodic boundary conditions. Initial configurations were prepared using the PACKMOL \cite{Packmol_Martinez2009} package. The Nose--Hoover thermostat and barostat \cite{Thermo_Hoover1985,Thermo_Nose2002} with damping parameters set to 0.1~ps and 1~ps, respectively, were used to maintain a constant temperature and a pressure of 0~bar. Properties were calculated in simulation cells containing 800 and 832 atoms for the liquid alloy and the molten salt, respectively. Prior to the production run, the systems were equilibrated for 50~ps. 

Densities of the systems were calculated in the NPT ensemble. The total number of frames analyzed for every composition and temperature was 500; these were evenly sampled from 500~ps (for the liquid alloy) and 1~ns (for the molten salt) long trajectories. The local structure was analyzed via partial radial distribution functions (RDFs). These distribution functions were calculated from 100 configurations evenly sampled from 100 ps equilibrium MD trajectories. For the metallic melt, we employed larger 4000-atom systems for the RDF analysis to access longer-range distances. 

The specific heat capacity at constant pressure, $C_P$, was calculated by direct estimation of the slope of the temperature $T$ dependence of enthalpy $H$:
\begin{equation}
H(T) = C_P \cdot T + {\rm const}.
\label{Cp_direct}
\end{equation}
In Equation \ref{Cp_direct}, enthalpy $H$ is substituted by internal energy $E$ in our calculations, since the simulations were carried out at a pressure of 0~bar.

For the studied systems, the temperature dependencies of viscosity, diffusion coefficients, and ionic conductivity (for the molten salt system) were calculated using MTP-MD. We employed the same trajectories that were used for the density calculations.

For viscosity calculations, we employed the Green--Kubo (GK) theory \cite{GK_Green1954,GK_Kubo1957,GK_Hess2002,GK_Allen2017,GK_Maginn2018}, computing the integral of the pressure tensor autocorrelation function. The integral was truncated after 5 ps for the liquid alloy and 25 ps for the molten salt, based on our convergence tests of the autocorrelation decay. The diffusion coefficients for each atomic species were obtained from the slope of the mean-squared displacement in the diffusive regime, according to the Einstein relation \cite{GK_Maginn2018}. 

For the Ca-Cu molten alloy, we tested the Stokes--Einstein--Sutherland (SES) equation
\begin{equation}
D = \frac{k_B T}{b \pi \eta R_h},
\label{SES}
\end{equation}
where $D$ is the self-diffusion coefficient of the particle, $k_B$ is the Boltzmann constant, $T$ is the temperature in the system, $\eta$ is the viscosity of the system, $R_h$ is the hydrodynamic radius of the particle. As we calculated viscosity and diffusion coefficients independently in our methodology, we checked whether $b=4$ or $b=6$ is more appropriate.

For the molten salt we calculated the temperature dependence of ionic conductivity, $\sigma$. We employed the GK relation:
\begin{equation}
\sigma = \frac{1}{3V k_B T} \int_0^{\infty} dt \langle {\bf J}_q(t) \cdot {\bf J}_q(0)\rangle,
\label{ec_GK}
\end{equation}
where $V$ is the volume of the system, $k_B$ is the Boltzmann constant, $T$ is the temperature in the system. ${\bf J}_{q}(t)$ is the charge flux 
\begin{equation}
{\bf J}_q(t) = e\sum_{i=1}^{N} z_i {\bf v}_i(t),
\label{J}
\end{equation}
where $e$ is the elementary charge, $N$ counts all the ions in the system, $z_i$ and ${\bf v}_i$ are the charge and velocity of a particle. Ions were assigned fixed partial charges in our calculations: $z_{Ca}=+2$, $z_{K}=+1$, and $z_{Cl}=-1$. We truncated the GK integral at 25~ps. Additionally, for calculations of ionic conductivity we tested the Nernst--Einstein (NE) formula:
\begin{equation}
\sigma = \frac{e^2}{Vk_BT}\sum_{k}N_{k}z^2_{k}D_{k},
\label{NE}
\end{equation}
where $e$ is the electron charge, $N_{k}$ is the number of $k$ species ions, $z_{k}$ is the charge of $k$ species ions, $D_{k}$ is the diffusion coefficient of $k$ species ions. The NE formula is an approximation, which is valid in the absence of correlations in ionic motion. We compared the approach accounting for all correlations (the GK method) with the NE formalism. 

Thermal conductivity was computed using the non-equilibrium Muller-Plathe method \cite{MP_method1997}, which imposes a heat flux and measures the resulting temperature gradient via Fourier's law. We used a $3.6~\text{nm} \times 3.6~\text{nm} \times 18~\text{nm}$ cell for the simulations, with a temperature gradient established along the z-axis. After a 500~ps equilibration, data were collected over a 2000~ps production run. The thermal conductivity was computed only for the molten salt. We omitted this property for the Ca-Cu alloy because the ionic contribution accounted for by classical MD is negligible compared to the electronic contribution. The latter requires additional DFT calculations beyond the scope of this work \cite{EC_knyazev2019,EC_galtsov2024}.

\subsection{Experimental measurements}
\label{subsec:experimental}

To validate the molecular dynamics calculations, several properties of the CaCl$_2$-KCl molten salt relevant to electrolysis were measured experimentally. The temperature dependencies of density, viscosity, and ionic conductivity were obtained. Details of the experimental techniques and setups are provided in Section 2 of the Supplementary Information.

\section{Results and discussion}
\label{sec:results}
\subsection{Potential fitting}

Using the strategies described in Section \ref{sec:methods}, we successfully developed MTPs for both the Ca-Cu liquid alloy and the CaCl$_2$-KCl molten salt. The resulting training and validation set sizes are reported in Table~\ref{table:errors}, along with the errors in energies per atom and forces acting on atoms. Correlation plots between the DFT and MTP are presented in Section 3 of the Supplementary Information for both systems.

For both systems, we obtained errors of approximately 5~meV/atom in energies per atom. The forces in atomic configurations are also reproduced with RMSEs of 80 and 136~meV/\si{\angstrom} for the Ca-Cu alloy and the CaCl$_2$-KCl electrolyte, respectively. These low errors confirm that the MTPs accurately reproduce the DFT data. Moreover, for the liquid alloy system, the potential is able to describe interactions in configurations spanning the entire range of compositions. This establishes a basis for the MTP for Ca-Cu to be compositionally transferable. 

Since the obtained potentials reproduce the interactions accurately, we expect the structural and dynamic properties of the liquids to be well-reproduced. A force error on the order of 100~meV/\si{\angstrom} has been shown to be sufficient for reproducing of the properties of atomic systems \cite{errors_Feng2022,errors_Kamaeva2024,errors_Rybin2024}. This provides a reliable foundation for calculating physicochemical properties.

\begin{table}[h!]
\caption{Summary on the trained potentials and their errors in validation. N$_{train}$ -- number of configurations in the training set. N$_{valid}$ -- number of configurations in the validation set. RMSE is reported as the error of potential for both energies and forces.}
\centering
\begin{tabular}{|c|c|c|c|c|c|}
\hline
System & Level & N$_{train}$ & N$_{valid}$ & \begin{tabular}[c]{@{}c@{}}Energy error, \\ meV/atom\end{tabular} & \begin{tabular}[c]{@{}c@{}}Force error,\\ meV/\si{\angstrom}\end{tabular} \\ \hline
Ca-Cu & 12 & 463 & 240 & 5.4 & 80 \\ \hline
CaCl$_2$-KCl & 14 & 591 & 200 & 5.0 & 136 \\ \hline
\end{tabular}
\label{table:errors}
\end{table}

\subsection{Physicochemical properties of Ca-Cu liquid alloy}

We benchmarked our potential against the structural properties of the melt. Density was calculated for temperatures ranging from 900 to 1400~K. At 1400~K, all the compositions are liquid. As the temperature decreases, the range of liquid compositions diminishes \cite{bruzzone1971}. Therefore, in accordance with the phase diagram, we calculated the density of the alloy only for the liquid phase. 

\begin{figure}[h!]
	\centering
	\includegraphics[trim={0cm 0cm 0cm 0cm}, clip, width=1\linewidth]{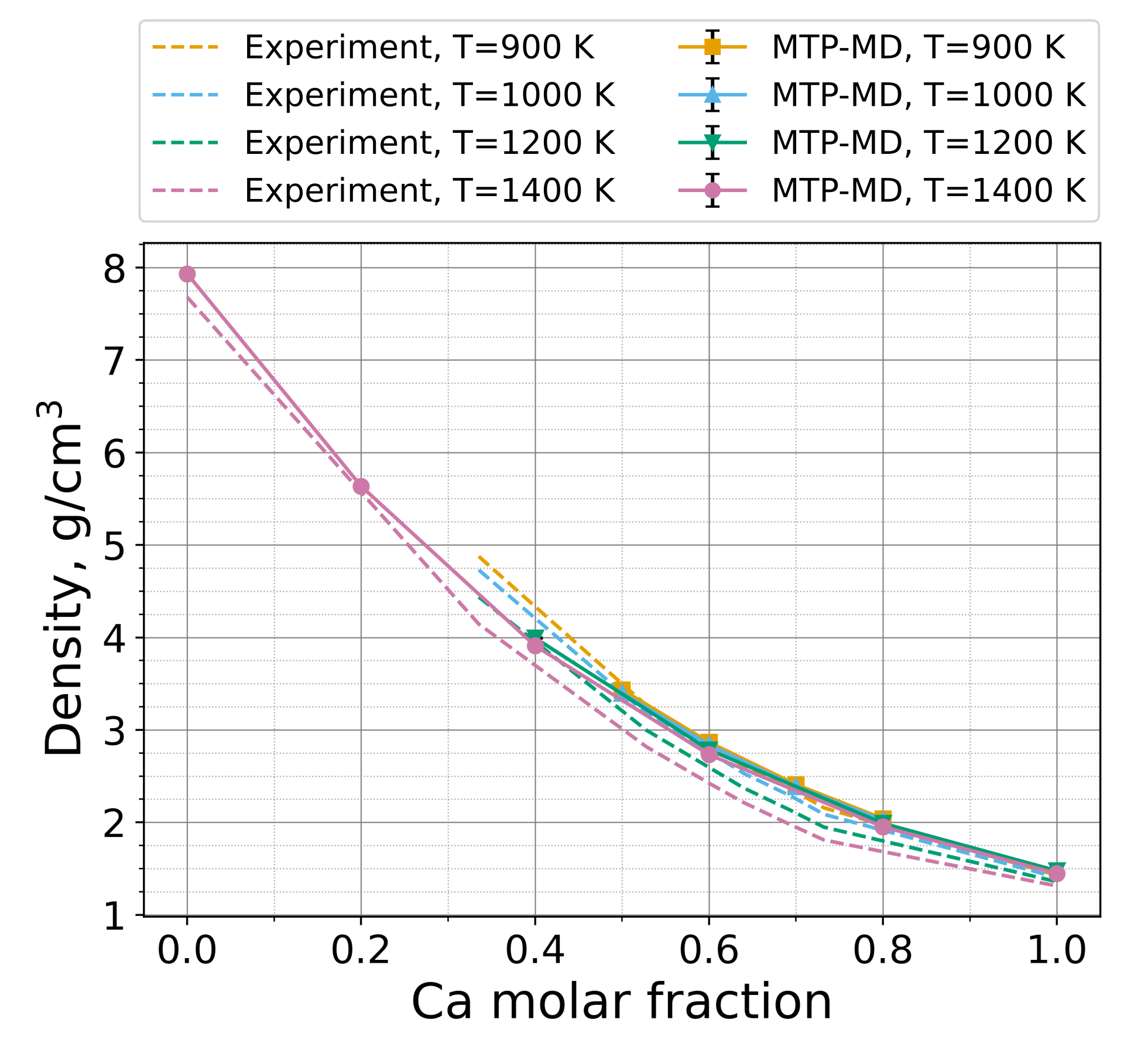}
 	\caption{Composition dependence of Ca-Cu liquid alloy density at temperatures 900, 1000, 1200, and 1400~K. Experimental data for Ca-Cu liquid alloy are taken from \cite{calcium_Hiemstra1997,CaCu_exp_Zaikov,copper_Assael2010}.}
\label{fig:cacu_dens}
\end{figure}

In Figure \ref{fig:cacu_dens}, we present the results of our MTP-MD calculations. The potential reproduces the density at lower temperatures (900 and 1000~K) with good accuracy. For alloys with a calcium molar fraction from 0.5 to 0.8, the relative errors compared to literature data \cite{CaCu_exp_Zaikov} are approximately 3--4\%. The potential also reproduces the density well for pure copper. For liquid copper at $T$=1400~K we obtained a density of 7.93~g/cm$^3$, while the experimental value is 7.68~g/cm$^3$ (a relative error of 3\%) \cite{copper_Assael2010}. However, for liquid calcium, the error is slightly higher. At a temperature of 1400~K, the experimental density is 1.31~g/cm$^3$ \cite{calcium_Hiemstra1997}, while the MTP-MD result is 1.44~g/cm$^3$ (a relative error of 10\%). The MTP-MD underestimates the thermal expansion of Ca-Cu liquid alloy: at higher temperatures, the simulation overestimates the density. However, the density estimation for the temperature range relevant to electrolysis is accurate. 

\begin{figure}[h!]
	\centering
	\includegraphics[trim={0cm 0cm 0cm 0cm}, clip, width=1\linewidth]{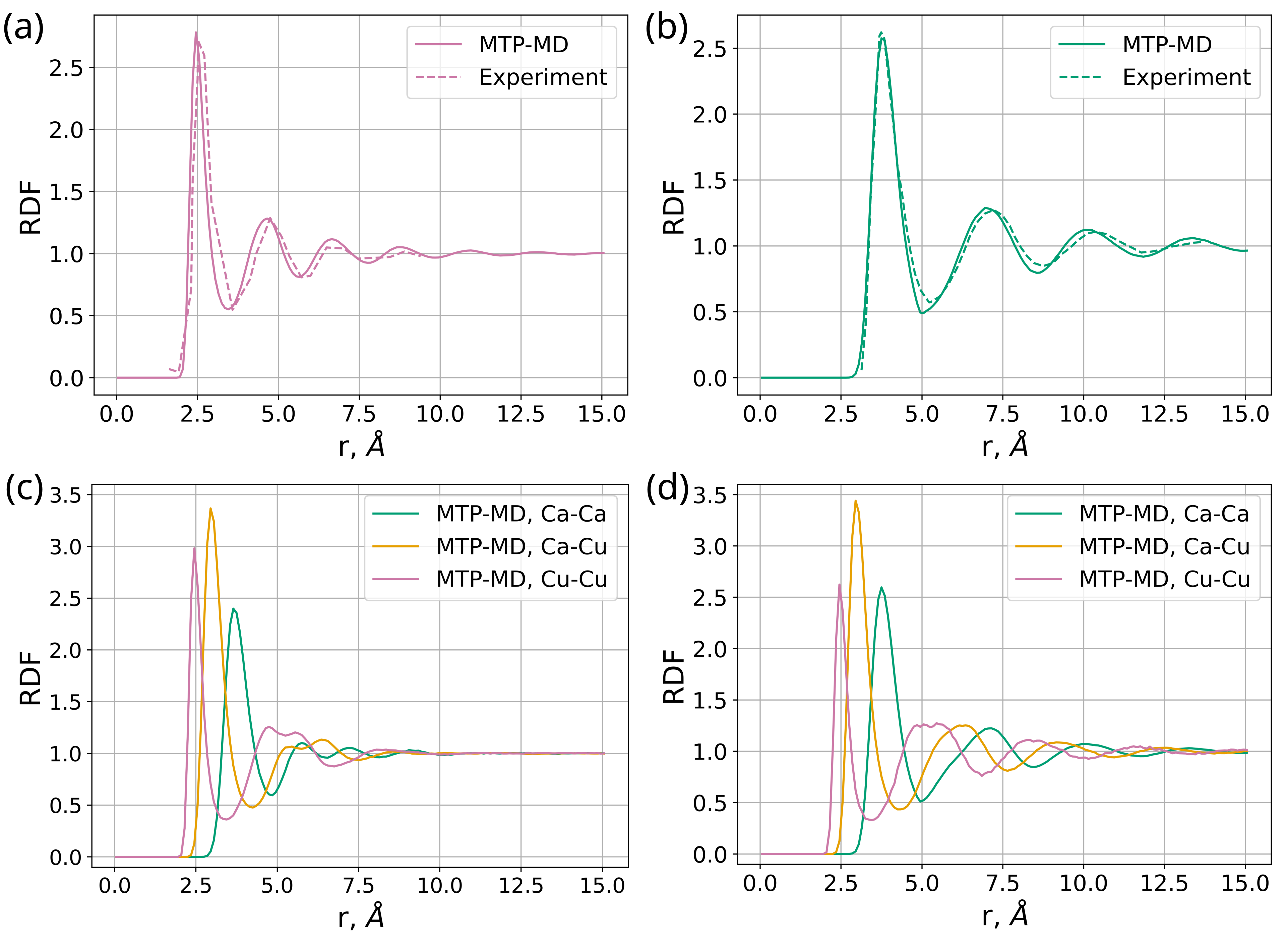}
 	\caption{RDFs for liquid alloys systems. (a) RDF in pure copper at temperature 1423~K. Experimental curve from \cite{CaRDF_Waseda1974}. (b) RDF in pure calcium at temperature 1123~K. Experimental curve from \cite{waseda1980structure}. (c-d) RDFs in Ca-Cu liquid alloy at molar fraction of calcium 0.5 and 0.8, temperature 1000~K. }
\label{fig:cacu_gofr}
\end{figure}

In Figures \ref{fig:cacu_gofr} (a) and (b), we report RDFs for liquid calcium (at 1123~K) and copper (at 1423~K) to provide additional validation of the potential, comparing RDFs to literature data \cite{CaRDF_Waseda1974,CaRDF_Baria2010,waseda1980structure}. For the RDFs of the pure alloy melts, excellent agreement is observed between the positions of the peaks and their heights. Thus, we conclude that the MTP captures the short-range structure of pure liquid metals. In Figures \ref{fig:cacu_gofr} (c) and (d), we report the RDFs for melts with a calcium molar fraction of 0.5 and 0.8 for the first time. In both melts, we observe a structure with local density oscillations specific to liquids. In the melt with a higher concentration of calcium, the strength of the secondary peaks decays more slowly, and more peaks in RDFs can be distinguished. The atomic radii of the species are consistent with the first peak positions in the RDFs. 

Structural properties of the Ca-Cu melts are reproduced across a wide range of compositions. This provides strong evidence of the applicability of the MTP to liquid alloy systems.

\begin{figure}[h!]
	\centering
	\includegraphics[trim={0cm 0cm 0cm 0cm}, clip, width=1\linewidth]{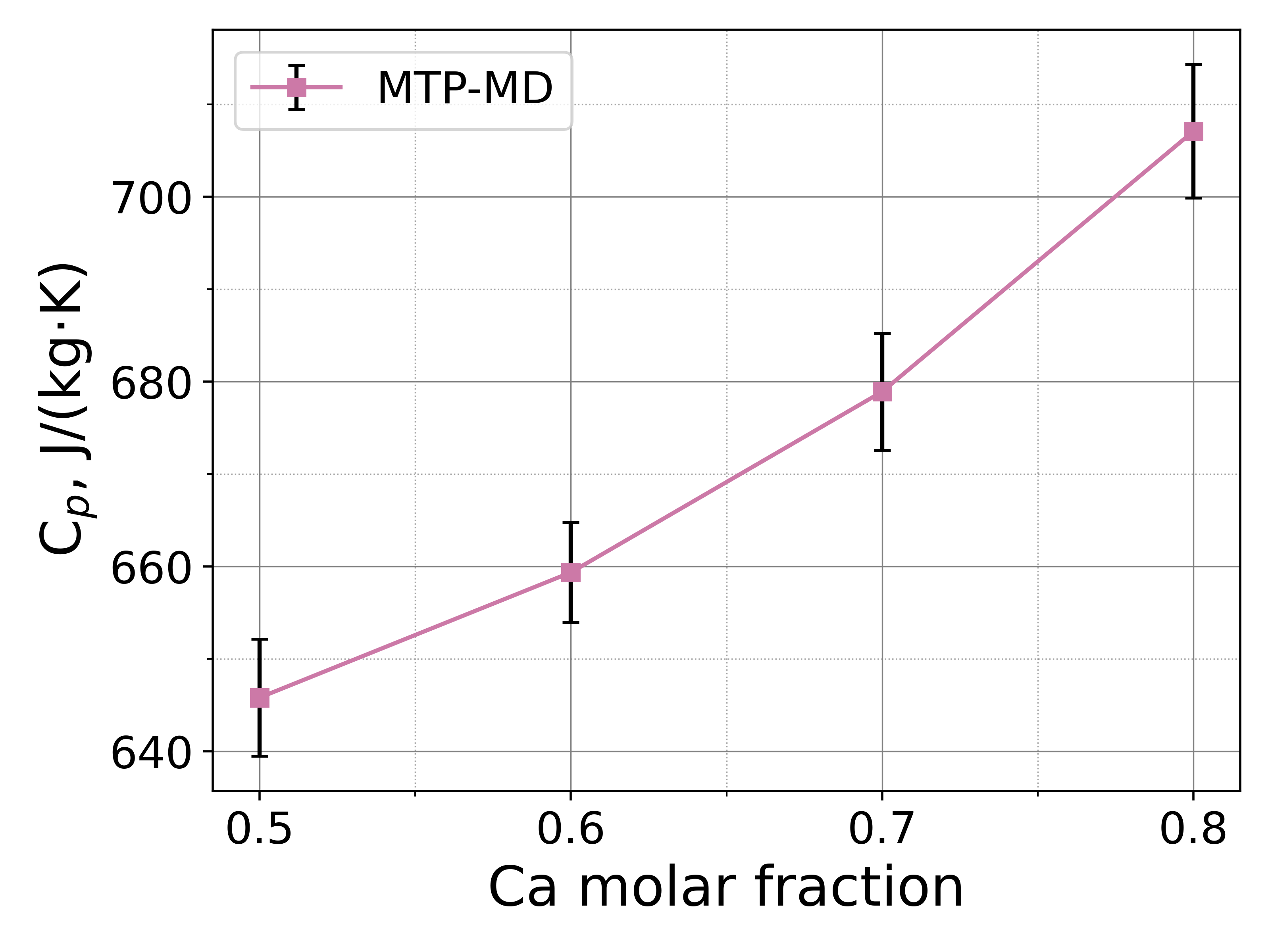}
 	\caption{Composition dependence of mass-specific heat capacity for Ca-Cu melt.}
\label{fig:cacu_SHC}
\end{figure}

The composition dependence of the specific heat capacity at constant pressure was calculated for the Ca-Cu melt. In Figure \ref{fig:cacu_SHC}, we report the calculated mass-specific heat capacity in J/(kg$\cdot$K). The calculations reveal that the mass-specific heat capacity increases nonlinearly with increasing calcium molar fraction. No experimental data are available for the mass-specific heat capacity of Ca-Cu liquid alloys. However, the values calculated with MTP-MD fall between the experimental values for the pure melts of calcium and copper. The experimental data for pure calcium range from 798 to 948~J/(kg$\cdot$K) \cite{CaThermExp_Abdullaev2023}. The mass-specific heat capacity of liquid copper is reported to be 571~J/(kg$\cdot$K) in Ref. \cite{Cu_SHC_Chekhovskoi2000}; in other studies, it ranges from 409 to 526~J/(kg$\cdot$K) \cite{Cu_SHC_Chekhovskoi2000}. The molar-specific heat capacity in J/(mol$\cdot$K) shows the opposite trend and decreases with increasing calcium concentration.

\begin{figure}[h!]
	\centering
	\includegraphics[trim={0cm 0cm 0cm 0cm}, clip, width=1\linewidth]{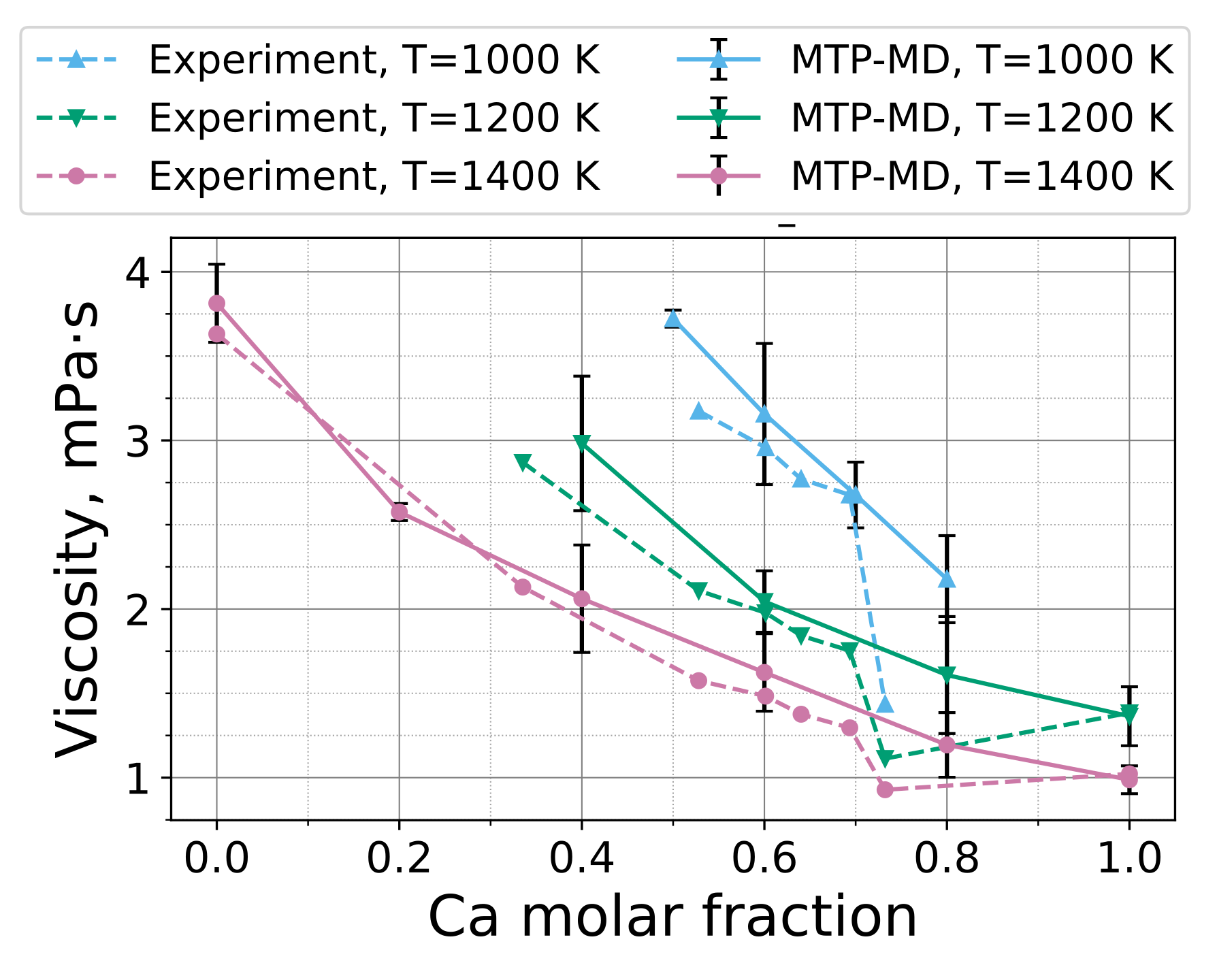}
 	\caption{Composition dependence of viscosity at temperatures 1000, 1200, and 1400~K. Experimental data for Ca-Cu liquid alloy is taken from \cite{CaCu_exp_Zaikov,copper_Assael2010}.}
\label{fig:cacu_visc}
\end{figure}

\begin{figure}[h!]
	\centering
	\includegraphics[trim={0cm 0cm 0cm 0cm}, clip, width=1\linewidth]{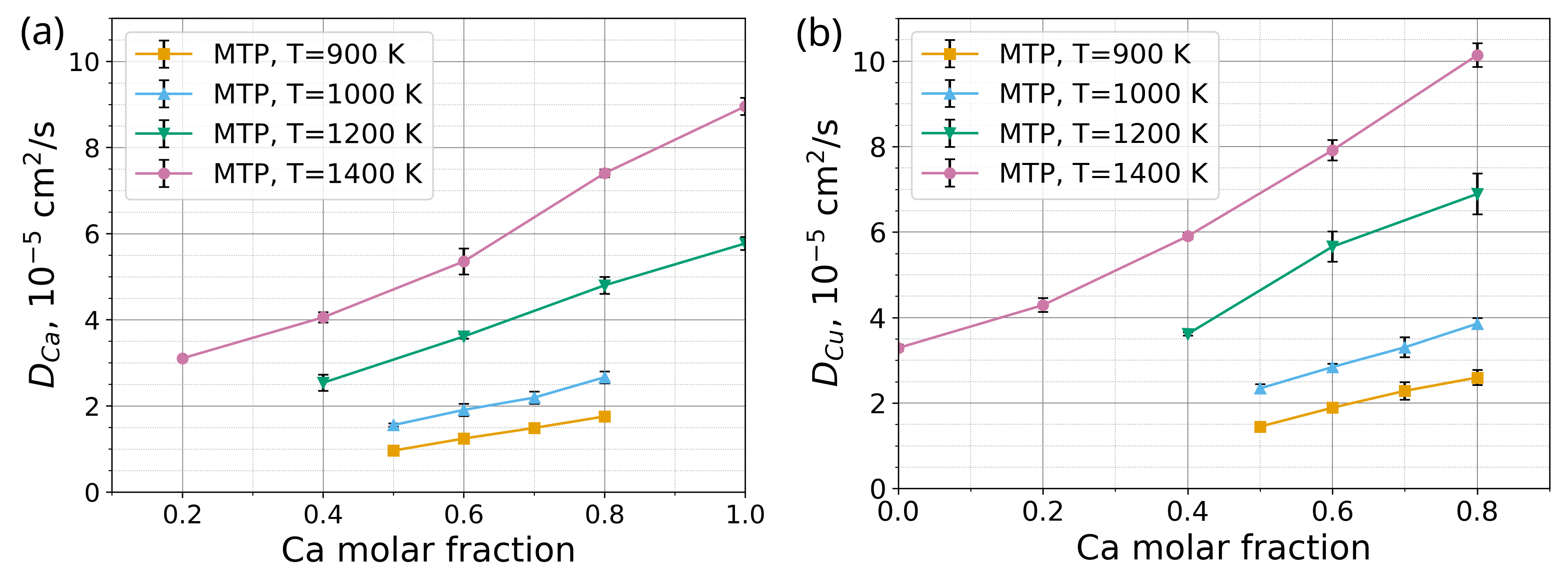}
 	\caption{Diffusion coefficients of Ca-Cu liquid alloy at temperatures 900, 1000, 1200, and 1400~K. (a) Composition dependence of calcium diffusion coefficient. (b) Composition dependence of copper diffusion coefficient.}
\label{fig:cacu_D}
\end{figure}

Figure \ref{fig:cacu_visc} presents the composition dependence of viscosity, comparing our MTP-MD results with literature data from \cite{CaCu_exp_Zaikov,copper_Assael2010}. Overall, the calculations show excellent agreement across all studied compositions. The viscosity exhibits an almost linear decrease with increasing calcium molar fraction. Notably, deviations from the linear trend in the literature data are most likely outliers. Thus, our model helps to resolve a physical inconsistency in the existing data. The accurate reproduction of the melt's viscosity provides confidence in the predicted diffusion coefficients, which are shown for calcium and copper in Figure \ref{fig:cacu_D}. The coefficients increase with higher calcium concentration, a trend that is consistent with the observed decrease in viscosity. Both dependencies are approximately linear. 

To test the SES equation \ref{SES}, we used the collected data. Since we obtained the diffusion coefficient and viscosity independently, we directly calculated the average value of $b$ for each system. For the Ca-Cu liquid alloy, we obtained a value of 3.7 $\pm$ 0.4. Slip boundary conditions ($b$=4) are more appropriate for these alloys than the widely used stick conditions ($b$=6). This conclusion is consistent with findings for other liquid alloys \cite{Ni_SES_Cherne2001,Ga_SES_Sheppard2015}. This finding is significant because the SES relation is widely used to estimate diffusion coefficients from viscosity data. Consequently, our results provide a refined value of $b$ that can be used to correct the diffusion coefficients derived from the viscosity data in \cite{CaCu_exp_Zaikov} and for further calculations in other alloys.

The agreement of the calculated properties of the Ca-Cu liquid alloy demonstrates that we have successfully developed a machine-learned potential that is reliable and compositionally transferable. Its accuracy across this range of structural, thermodynamic, and transport properties confirms the MTP's robustness for simulating liquid alloy systems.

\subsection{Physicochemical properties of the CaCl$_2$-KCl molten salt}

\begin{figure}[h!]
	\centering
	\includegraphics[trim={0cm 0cm 0cm 0cm}, clip, width=1\linewidth]{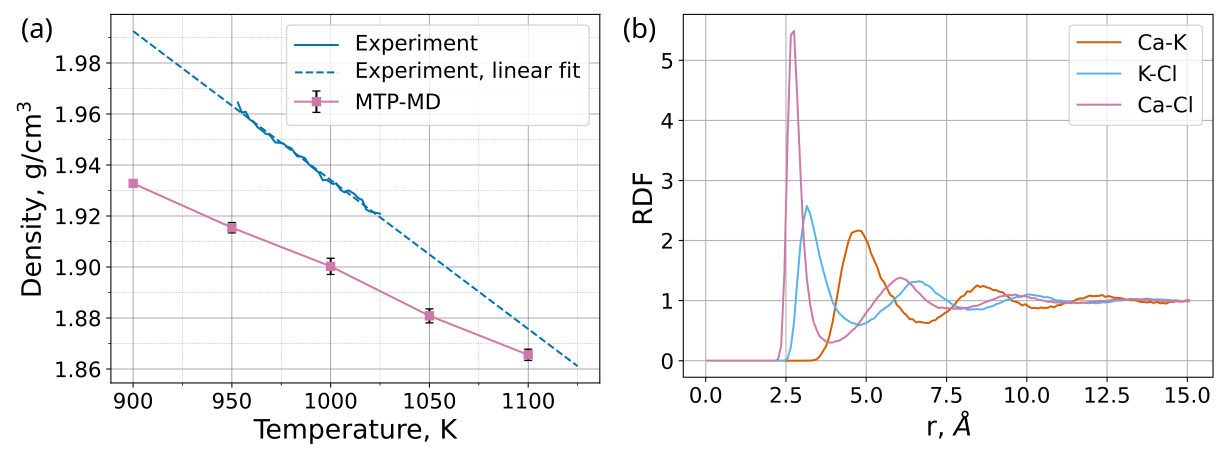}
 	\caption{(a) Temperature dependence of the CaCl$_2$-KCl molten salt density. MTP-MD results (pink squares) are compared with our experimental data (blue line). (b) CaCl$_2$-KCl molten salt RDFs calculated at temperature 1000~K.}
\label{fig:cakcl_str}
\end{figure}

Next, we tested the calculation of physicochemical properties using the MTP prepared for the molten salt system. The density of the electrolyte was calculated over the temperature range from 900 to 1100~K. In Figure \ref{fig:cakcl_str} (a), we report the temperature dependence of the density at a fixed composition. Our calculations are in good agreement with the experimental measurements conducted in this work, with a relative error of approximately 2--3\%. This quantitative validation confirms the reliability of our approach using MD with the MTP for the structural properties of molten salts. 

In the RDFs presented in Figure \ref{fig:cakcl_str} (b), calculated at a temperature of 1000~K, we observe that calcium coordinates with chlorine atoms more strongly than potassium. The Ca--Cl peak is sharper and more intense than the K--Cl peak, which indicates stronger and more localized bonding between calcium and chlorine. The positions of the peaks are consistent with the atomic radii.

The MTP-MD specific heat capacity of the molten salt is 980$\pm$10~J/(kg$\cdot$K), which agrees well with the experimental value of approximately 960~J/(kg$\cdot$K) reported for similar compositions \cite{physical_janz1979}. At 1000 K, the thermal conductivity was determined to be 0.438~W/(m$\cdot$K). This result is consistent with a previous computational study that employed the more expensive DeepMD potential \cite{DeepMD_CaKCl_Bu2021} and with the experimental value of 0.436~W/(m$\cdot$K) for the pure CaCl$_2$ molten salt at 1000~K \cite{pure_GHERIBI2014}. These agreements demonstrate that the MTP-MD accurately reproduces the thermodynamic properties and heat transport phenomena in the CaCl$_2$-KCl system.

\begin{figure}[h!]
	\centering
	\includegraphics[trim={0cm 0cm 0cm 0cm}, clip, width=1\linewidth]{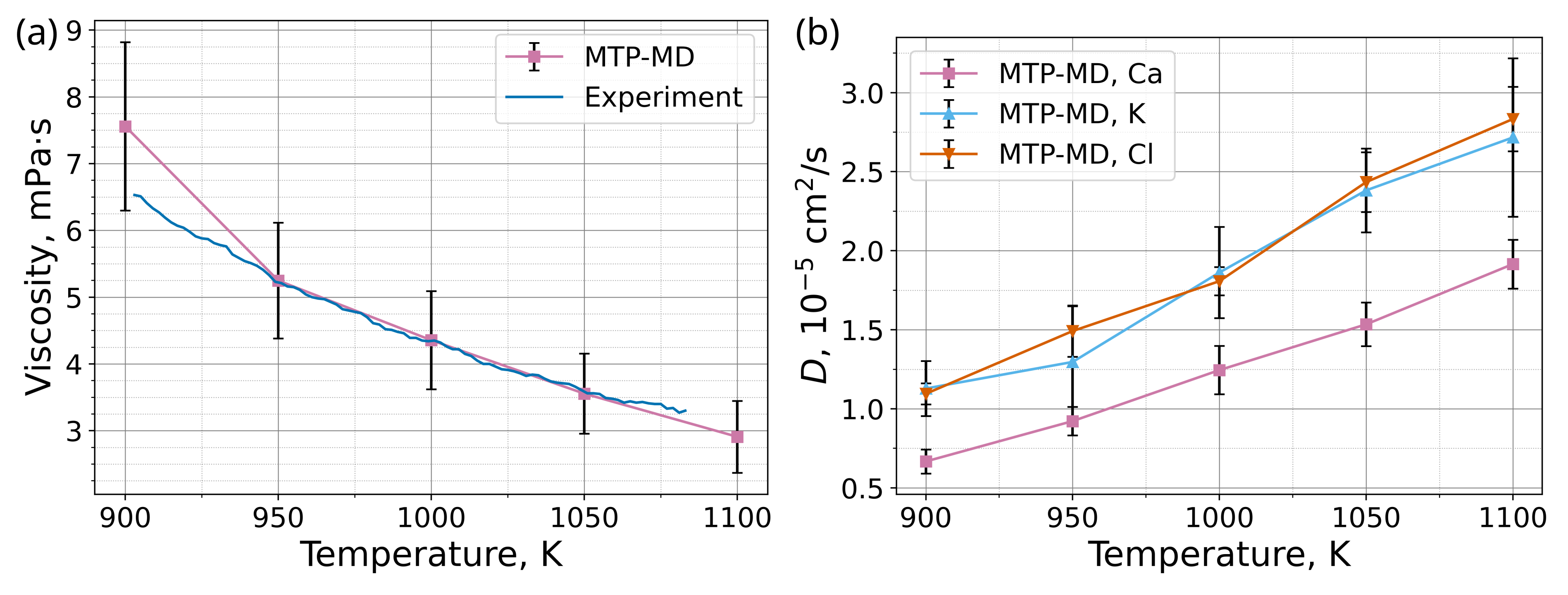}
 	\caption{Transport properties of the CaCl$_2$-KCl molten salt in the temperature range of 900-1100~K. (a) Viscosity as a function of temperature. Pink circles show the MTP-MD results, while the blue line represents our experimental data. (b) Diffusion coefficients as a function of temperature for Ca$^{2+}$ (pink squares), K$^+$ (blue triangles), and Cl$^-$ (orange triangles) ions.}
\label{fig:cakcl_transport}
\end{figure}

Next, we calculated the transport properties of the CaCl$_2$-KCl molten salt, which are key for electrolysis efficiency: viscosity, diffusion coefficients, and ionic conductivity. In Figures \ref{fig:cakcl_transport}, we report the viscosity (a) and diffusion coefficients of atoms in the molten salt (b). The viscosity of the electrolyte decreases with rising temperature, as expected. We observe excellent agreement between the MTP-MD calculations and the experimental data for viscosity, which were obtained for this study. The MTP-MD demonstrates an ability to capture subtle physicochemical trends in molten salts. The achieved agreement for viscosity provides a solid foundation for reliable diffusion coefficient calculations. In the temperature dependence of the diffusion coefficients, we observe that calcium atoms are less mobile than potassium. While potassium has a larger atomic radius and would be expected to be less mobile, our result indicates that in CaCl$_2$-KCl, calcium ions coordinate with chlorine atoms and consequently have a larger effective hydrodynamic radius.

\begin{figure}[h!]
	\centering
	\includegraphics[trim={0cm 0cm 0cm 0cm}, clip, width=1\linewidth]{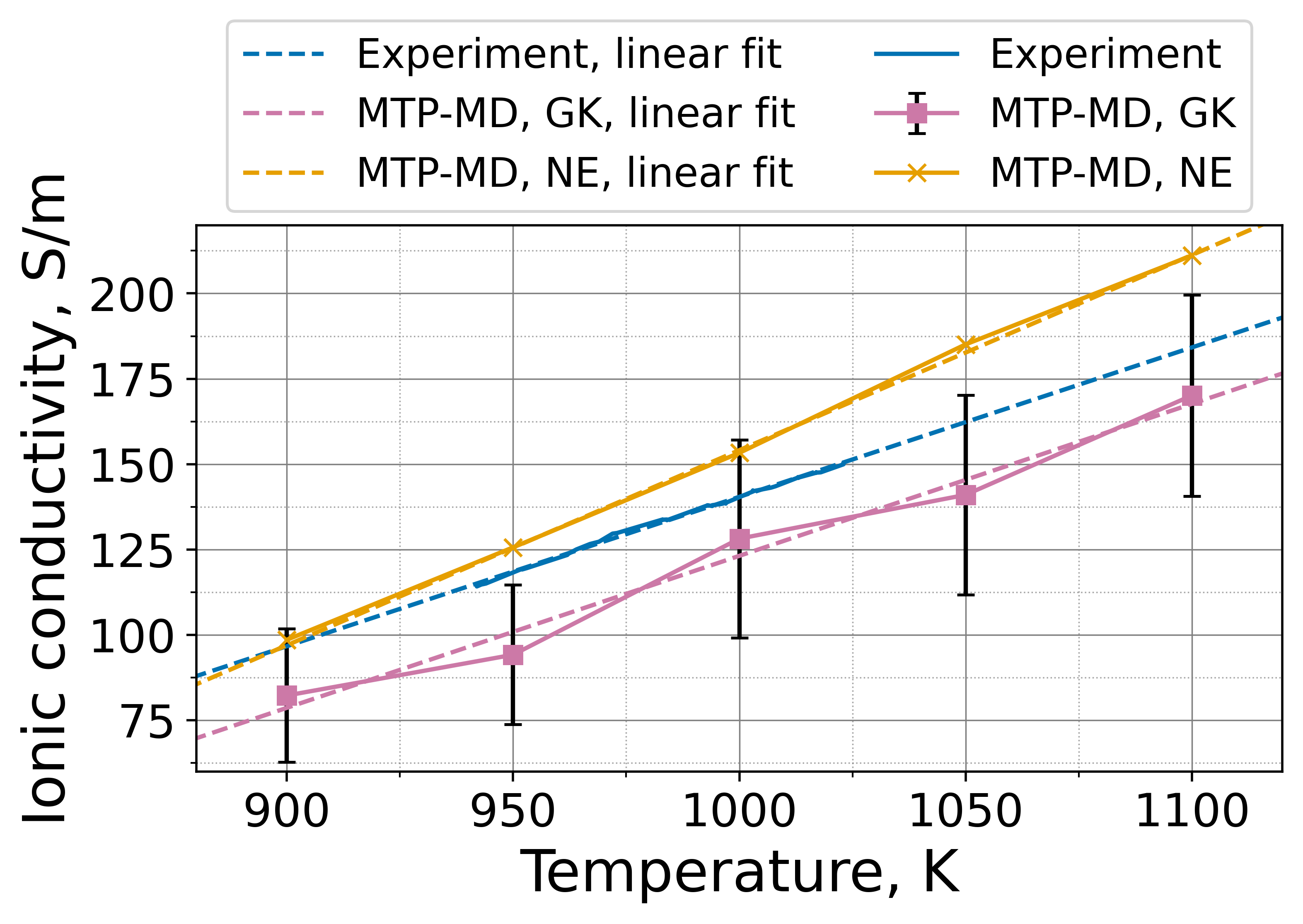}
 	\caption{Temperature dependence of ionic conductivity in the CaCl$_2$-KCl molten salt. Results from the Green-Kubo (GK, pink squares) and Nernst-Einstein (NE, golden crosses) methods are compared with experimental data (blue line) obtained in this work. Solid lines represent the calculated or experimental data, while dashed lines show linear fits.}
\label{fig:cakcl_sigma}
\end{figure}

As shown in Figure \ref{fig:cakcl_sigma}, the ionic conductivity exhibits a linear increase with temperature, a trend captured by both the GK and NE formalisms. This correctly reproduces the expected thermodynamic behavior, which reflects enhanced ion mobility at higher temperatures (see Figure \ref{fig:cakcl_transport}).

The GK method consistently underestimates the ionic conductivity by 10--25~S/m, resulting in a relative error of approximately 6--20\%. However, with the respect to uncertainty of calculations, the agreement between the calculations and experimental measurements are reached. Relative errors are lower for the NE method, which are 4--15\% over the measured temperatures; however, the NE approach predicts an incorrect slope for the temperature dependence of ionic conductivity, as shown by the fit in Figure \ref{fig:cakcl_sigma}. This leads to larger errors in the high-temperature region, where the experimental data are extrapolated. The GK method does not have this disadvantage. Moreover, the data on diffusion coefficients from Ref. \cite{DeepMD_CaKCl_Bu2021} indicate that at higher calcium concentrations, the atomic coordination is higher, which will lead to larger errors for the NE method. Therefore, we still recommend using the method that accounts for all the correlations in ionic motion. In Section 4 of Supplementary Information the additional details on the conductivity calculations using the GK method are provided. The underestimation of ionic conductivity by GK can be attributed to the system size dependence of ionic conductivity, which will be studied extensively in future research. 

In summary, the MTP-MD approach provides a robust framework for predicting properties of molten salts. It accurately reproduces densities, viscosities, thermal conductivities, and ionic conductivities, along with their correct thermodynamic trends. 

\clearpage
\section{Conclusion}
\label{sec:conclusion}

In this work, we demonstrate that MTP-MD provides an accurate and efficient framework for simulating key systems in calcium electrolysis. Applied to the molten Ca-Cu cathode, the MTP's compositional transferability allows a single potential to reproduce physicochemical properties across the entire liquid composition range. For the CaCl$_2$-KCl molten salt electrolyte, we show that a local MTP effectively captures the structure and dynamics of the ionic melt. The results indicate that the MTP implicitly handles short-range electrostatics, enabling stable simulations. Furthermore, MTP-MD is significantly faster than AIMD.

An additional advantage of our approach is the use of MTP-MD to compute a wide range of physicochemical properties without fitting to experimental data. The calculated densities and radial distribution functions agree well with available experimental data, with errors of 3--4\% in the temperature range relevant to electrolysis. We report specific heat capacities for both systems that are in quantitative agreement with the existing data. MTP-MD accurately predicts transport properties, including viscosity and ionic conductivity (with a maximal relative error of 20\%). This confirms that MTP-MD reliably captures both the structure and dynamics of these melts.

The success in modeling CaCl$_2$-KCl promises expansion to related electrolytes such as CaO-CaCl$_2$-KCl, CaCl$_2$-CaF$_2$, and CaCl$_2$-NaCl, enabling the prediction of how additives affect properties relevant to electrolysis. More broadly, the approach can be applied to optimize other high-temperature molten salt processes, e.g., production of aluminum, magnesium, and other metals.

Building on this validated foundation, the framework can be extended to tackle other critical challenges, such as calculating solubility and surface tension. Prior studies have successfully reported computed surface tension in alloys using MD \cite{ST_Juarez2023, ST_ML_Xiao2023}, and given the accuracy of our MTPs in the bulk phase, we anticipate no impediment to modeling vacuum--alloy interfaces. Calculating solubility, as attempted for molten salt reactor-relevant salts \cite{MAXWELL2022153633}, is a crucial next step, as it directly impacts electrolysis efficiency; for example, lower calcium solubility in the melt hinders Ca-Cu alloy enrichment and wastes energy. The calculation of these properties represents a clear and important direction for future work.

\section{Data availability}
\label{sec:data}

MTPs for the Ca-Cu and CaCl$_2$-KCl systems, along with the corresponding training and validation sets, are available in the \href{https://github.com/tregedron/CaCu-and-CaCl2-KCl-melts}{GitHub repository}. 

\section*{Author Contributions}
\label{sec:contributions}

Mikhail Polovinkin: methodology, investigation, validation, conceptualization, data curation, writing –- original draft. Nikita Rybin: conceptualization, methodology, supervision, writing –- review and editing. Dmitrii Maksimov: conceptualization. F. Valiev: investigation. A. Khudorozhkova: investigation. M. Laptev: investigation. A. Rudenko: investigation.  Alexander Shapeev: resources, funding acquisition, supervision.

\section*{Acknowledgments}
\label{sec:acknowledgments}

The work was supported by the grant for research centers in the field of AI provided by the Ministry of Economic Development of the Russian Federation in accordance with the agreement 000000C313925P4F0002 and the agreement with Skoltech №139-10-2025-033.

\bibliographystyle{elsarticle-num}

\clearpage
\bibliography{cas-refs}

\clearpage

\section*{Supplementary Information}

The Supplementary Information contains additional information and details of calculations supporting the main text. The experimental techniques used for the molten salt properties measurements are described. The justification of dDsC dispersion correction is presented. We compare DFT and MTP energy and forces acting on the atoms. Also, we provide two ionic conductivity GK integral vs time dependencies as the convergence test.

\section*{Dispersion correction selection.}
\label{appendix:diss_corr}
\begin{figure*}[h!]
	\centering
	\includegraphics[trim={0cm 0cm 0cm 0cm}, clip, width=1\linewidth]{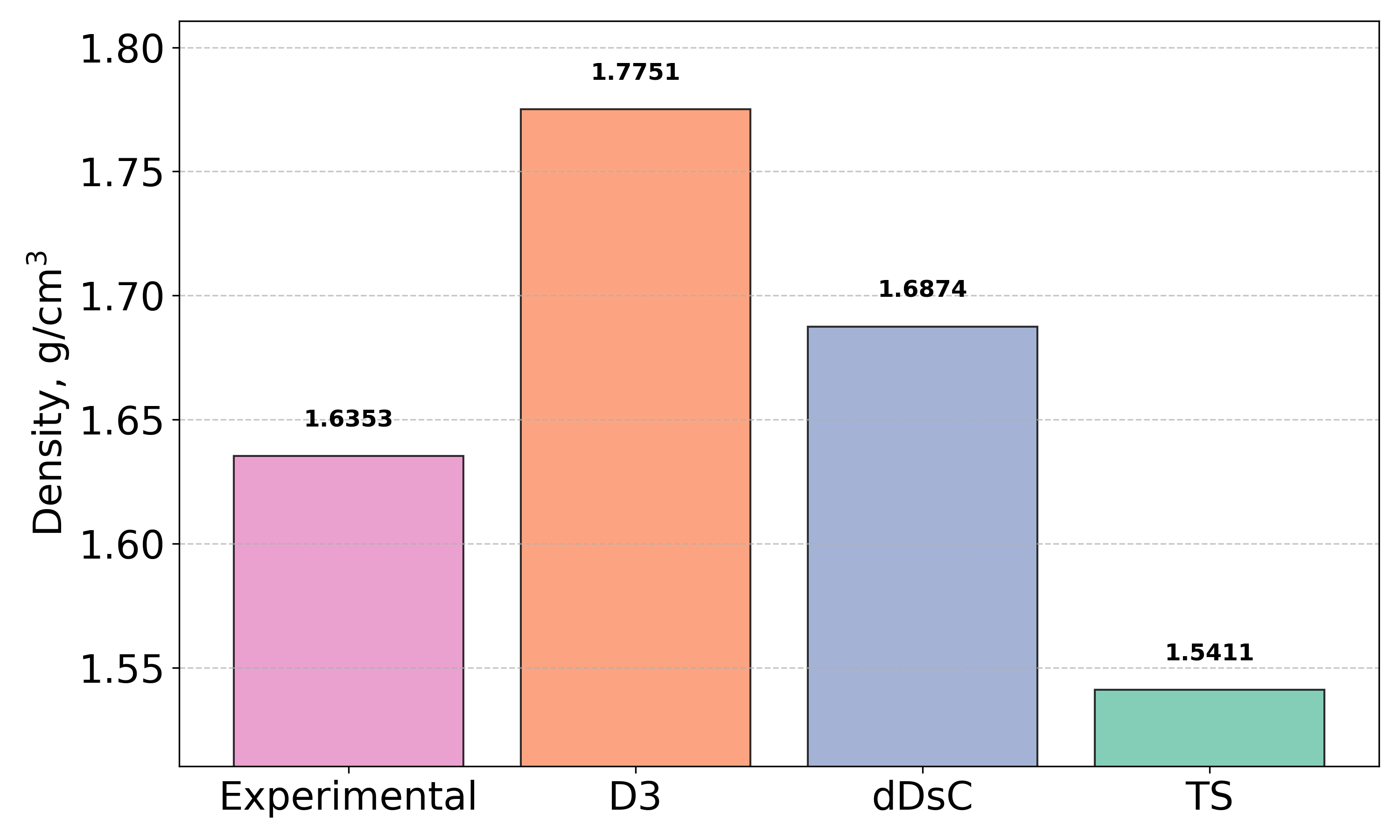}
 	\caption{Density of AlF$_3$-NaF molten salt calculated with potentials fitted on DFT with various dispersion corrections and experimental value \cite{AlFNaF}.}
\label{fig:density_correction}
\end{figure*}

In the AlF$_3$-NaF (1:1 molar fractions) molten salt at 973~K for in-house research we conducted experiments to choose the best performing dispersion correction in VASP. The D3 \cite{D3_Grimme2010,D3_Grimme2011}, TS \cite{TS_Tkatchenko2009} and dDsC \cite{dDsC_steinmann2011generalized,dDsC_steinmann2011} corrections were tested. The comparison of molten salt density for various dispersion corrections is presented in Figure \ref{fig:density_correction}.

\clearpage

\section*{Experimental methods}
\subsection*{Density measurements}

The density of molten salt systems was measured using the hydrostatic weighing method. The principle of the method is that a sinker of known volume, when immersed in a molten medium, displaces an equal volume of the liquid. Knowing the change in the sinker mass upon immersion in the melt and the sinker's volume, the density of the displaced liquid can be calculated using Equation (\ref{exp_density_1}):

\begin{equation}
    \rho = \frac{m_1 - m_2}{V}
    \label{exp_density_1}
\end{equation}

where $\rho$ is melt density, $m_1$ and $m_2$ mass of the sinker weighed in an argon atmosphere and in the melt correspondingly and $V$ is volume of the sinker.

\begin{figure*}[h!]
	\centering
	\includegraphics[trim={0cm 0cm 0cm 0cm}, clip, width=1\linewidth]{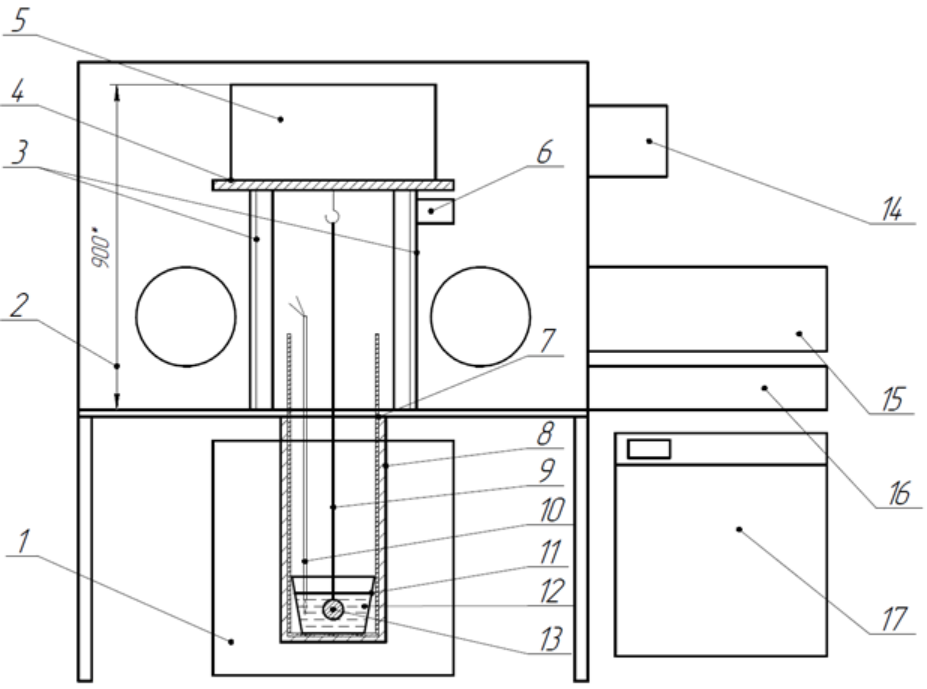}
\caption{Schematic of the glovebox-equipped measurement setup:
1 – Vertical split-tube resistance furnace of shaft type;
2 – Lifting-lowering mechanism;
3 – Vertical displacement unit (frame);
4 – Platform;
5 – Analytical balance with bottom suspension;
6 – Stepper motor;
7 – Quartz container;
8 – Retort;
9 – Platinum wire;
10 – Pt-Pt/Rh thermocouple (Type S) in alumina sheath;
11 – Glassy carbon crucible;
12 – Molten electrolyte under study;
13 – Platinum sinker;
14 – Electronic control unit;
15 – Glovebox main airlock;
16 – Glovebox small airlock;
17 – Glovebox gas purification unit.
}

\label{fig:density_setup_cacl}
\end{figure*}

The measurements were conducted in a glovebox setup under a high-purity argon atmosphere with an oxygen content of $< 0.1$~ppm (Figure \ref{fig:density_setup_cacl}). The measuring cell consisted of a quartz retort. A spherical platinum sinker was suspended on a platinum wire approximately 0.6~m long and 0.5~mm in diameter, connected to a Radwag XA210.4Y electronic balance of special (Class I) accuracy. The immersion and extraction of the platinum sinker into/from the melt were performed using a lifting mechanism. The sinker was weighed first in the gas atmosphere and then in the molten salt under study.

Before starting the measurements, the measuring cell was calibrated by determining the temperature dependence of the platinum sinker's volume. The sphere was immersed in a melt of known density, 0.51:0.49 NaCl–KCl (molar fractions). The melt density at each temperature was calculated using Equation (2) from reference \cite{janz1988thermodynamic}:
\begin{equation}
    \rho=2.1314-5.6793  \cdot 10^{-4} \cdot T(K)
\end{equation}

Using the mass difference at each temperature ($m_1-m_2$) and the calculated density $\rho_\text{calc}$ of the KCl-NaCl mixture, the sinker's volume was determined using Equation \ref{exp_density_2}:
\begin{equation}
    V = \frac{m_1 - m_2}{\rho_{\text{calc}}}
    \label{exp_density_2}
\end{equation}

where $\rho_\text{calc}$ is the density of the KCl-NaCl mixture melt calculated using Equation (\ref{exp_density_1}), $m_1$ and $m_2$ mass of the sinker weighed in an argon atmosphere and in the KCl-NaCl melt correspondingly and $V$ is volume of the platinum sinker.

Once the temperature dependence of the platinum sinker's volume was established, the density of the studied melt was measured.

\subsection*{Ionic conductivity measurements}

The ionic conductivity was measured using electrochemical impedance spectroscopy with parallel metallic (platinum) electrodes.

The method is based on recording the impedance of an electrochemical system containing the melt of the analyzed salt over an alternating current frequency range from 100~Hz to 150~kHz with an AC voltage amplitude of 10~mV. Measurements were performed using a Z-1500J impedance analyzer. The resistance of the molten electrolyte under study was determined from the impedance spectrum plot (the so-called Nyquist plot) by the real part of the impedance at the intersection point with the abscissa axis.

The resistance, determined by the frequency-independent active component of the impedance, is taken as the resistance of the molten electrolyte. This value is then used to calculate the ionic conductivity $\kappa$ ($S/m$)
of the analyzed electrolyte using Equation (\ref{exp_electro_1}):
\begin{equation}
    \kappa = \frac{K}{R_{\text{el}}}
    \label{exp_electro_1}
\end{equation}
where K is the cell constant and $R_\text{el}$ is the ohmic resistance of the electrolyte.
 
The temperature dependence of the cell constant was determined during calibration using parallel platinum electrodes and a reference salt melt with a known temperature-dependent conductivity. For this purpose, an equimolar KCl-NaCl mixture (0.5NaCl–0.5KCl) was used, the conductivity of which has been well-studied in the temperature range from 933~K to 1103~K.

\begin{figure*}[h!]
	\centering
	\includegraphics[trim={0cm 0cm 0cm 0cm}, clip, width=1\linewidth]{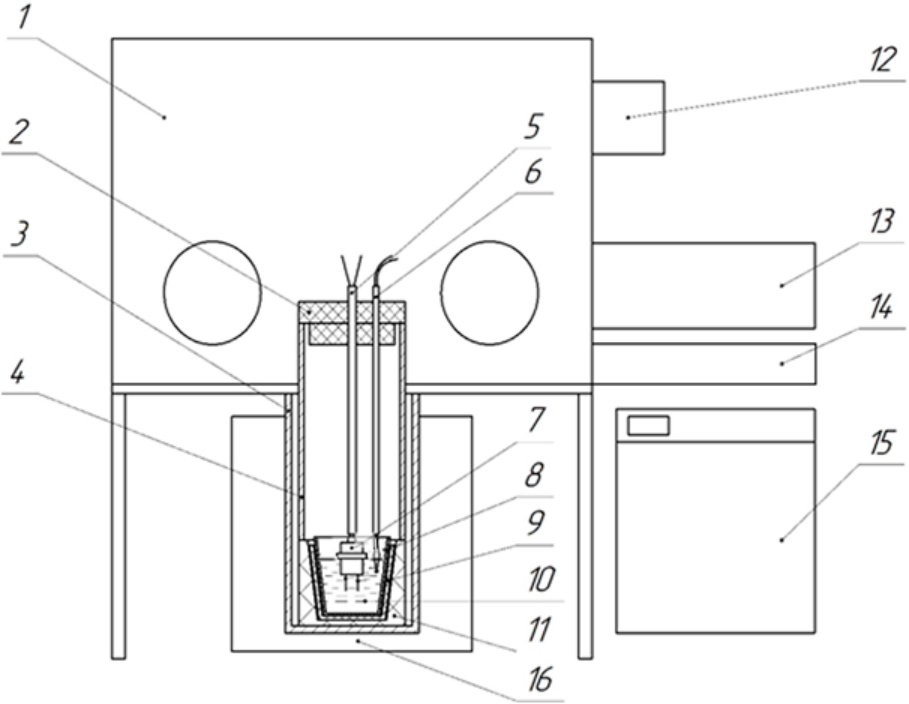}
\caption{Schematic diagram of the glovebox-equipped setup for specific conductivity measurements:
1 - Argon-atmosphere glovebox;
2 - Fluoroplastic cover;
3 - Retort;
4 - Stainless steel tube;
5 - Alumina electrode sheath;
6 - Pt-Rh/Pt thermocouple (Type S) in alumina sheath;
7 - Parallel platinum electrodes;
8 - Glassy carbon crucible;
9 - Alumina protection sleeve;
10 - Molten electrolyte under analysis;
11 - Graphite container;
12 - Electronic control unit;
13 - Glovebox main airlock;
14 - Glovebox small airlock;
15 - Glovebox gas purification unit;
16 - Vertical split-tube resistance furnace of shaft type.
}

\label{fig:el_cond_setup_cacl}
\end{figure*}

Both the cell calibration and the conductivity measurements of the molten electrolytes were performed using a glovebox-based measurement setup (Figure \ref{fig:el_cond_setup_cacl}).

Measurements were carried out at temperatures ranging from 953 to 1023~K. The ionic conductivity of the studied electrolytes at a given temperature (within the 953–1023~K range) was calculated using the cell constant determined at the same temperature.

\subsection*{Viscosity measurements}

The viscosity of molten salts was measured using a high-temperature rotational rheometer FRS-1600 (Anton Paar) in the temperature range of 573–1873~K. The operating principle involves the rotation of an inner graphite cylinder (diameter 15~mm, height 20~mm) within a stationary outer cylinder (inner diameter 30~mm), with a 2~mm gap filled with the molten sample. Viscosity measurements were conducted in an inert gas atmosphere, supplied through nozzles at the bottom of the setup. During heating, thermal expansion of the measurement system occurs, which is compensated by an automatic gap control system.

The sample mass (m, g) was calculated using the formula:
\begin{equation}
    m = v \rho
\end{equation}
where $v$ is the volume, 35$\pm$2~cm$^3$, and $\rho$ is the density of the molten sample at the measurement temperature.

The sample was loaded into the crucible in air. Heating was performed in a Carbolite STF16/180 furnace under an inert gas atmosphere. After melting (T$_\text{liquidus}$ = 914~K) and homogenization of the melt by rotor rotation for 40-50 minutes, the shear rate was measured. The rotational method is based on Newton’s law for the flow of a Newtonian fluid:

\begin{equation}
    \tau = \eta  \dot\gamma
\end{equation}

where $\tau$ is the shear stress, $\eta$ is the dynamic viscosity, and $\dot\gamma$ is the shear rate.

To ensure accurate viscosity measurements, laminar flow was established in the sample. The dependence of viscosity on shear rate was determined from flow and viscosity curves, and a shear rate of 11~s$^{-1}$ was selected, at which the melt behaves as a Newtonian fluid. Viscosity measurements were then conducted at a cooling rate of 2~K/min. The viscosity measurement methodology is described in detail in \cite{2023RuMet2023..141R}.

\clearpage

\section*{Correlation plots for energies and forces.}
\label{appendix:corr_plots}
\begin{figure*}[h!]
	\centering
	\includegraphics[trim={0cm 0cm 0cm 0cm}, clip, width=1\linewidth]{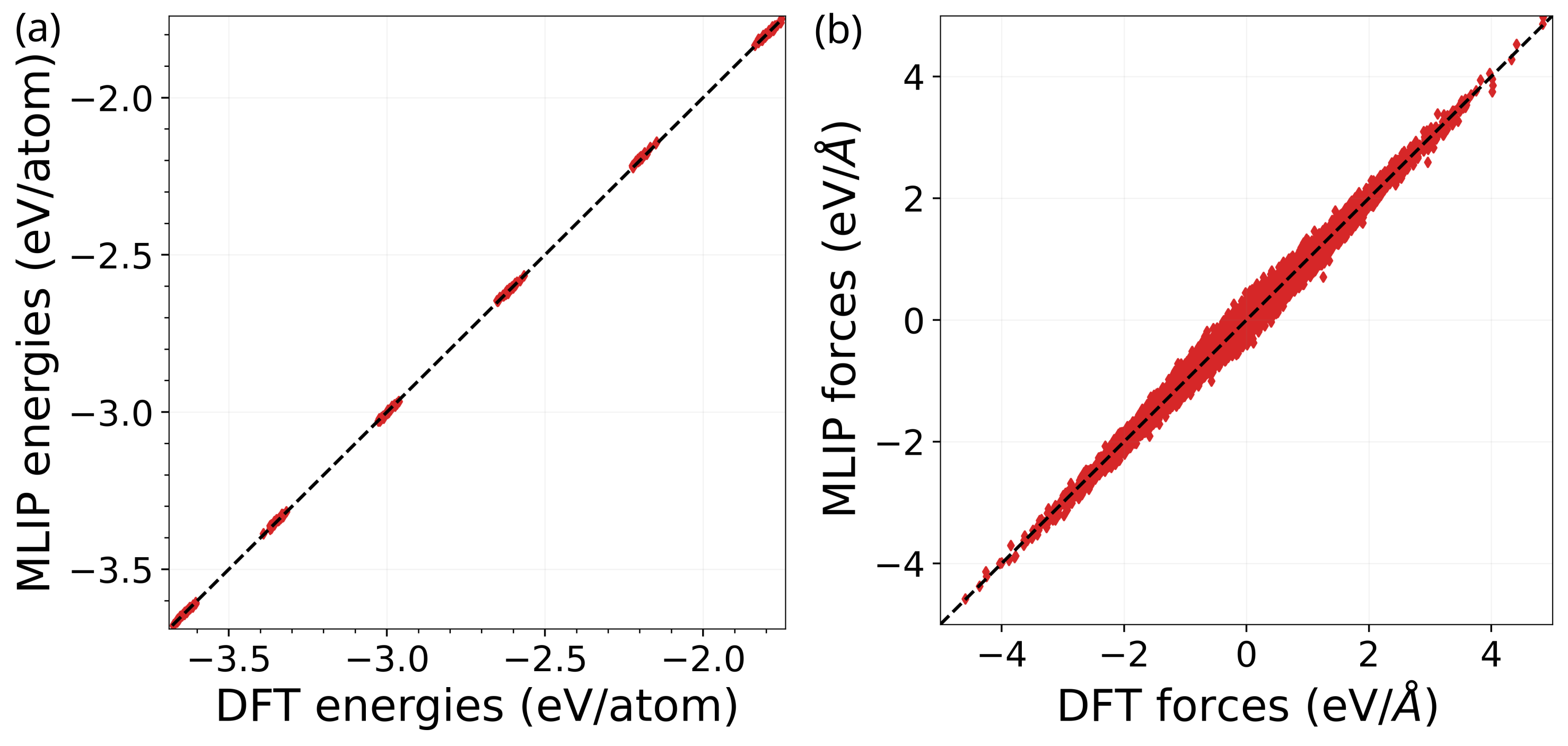}
 	\caption{Correlation plots of energies per atoms (a) and forces (b) in Ca-Cu validation set. The size of validation set was 240 configurations with calcium molar fraction from 0 to 1.}
\label{fig:errors_cacu}
\end{figure*}

\begin{figure*}[h!]
	\centering
	\includegraphics[trim={0cm 0cm 0cm 0cm}, clip, width=1\linewidth]{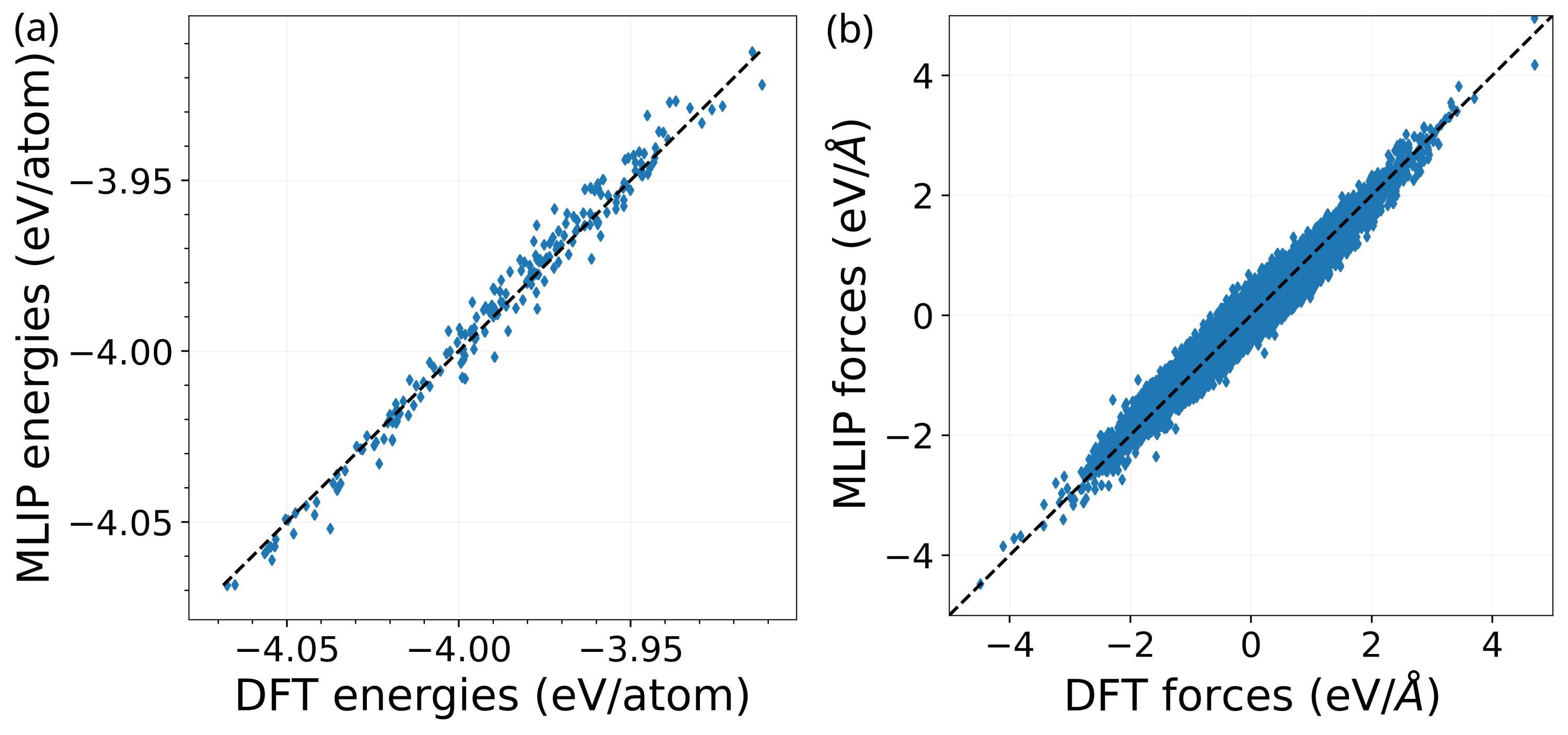}
 	\caption{Correlation plots of energies per atoms (a) and forces (b) in CaCl$_2$-KCl validation set. The size of validation set was 200 configurations.}
\label{fig:errors_cakcl}
\end{figure*}

In figures \ref{fig:errors_cacu} and \ref{fig:errors_cakcl} the correlation plots for energies and forces are presented for Ca-Cu and CaCl$_2$-KCl systems, respectively. These plots were obtained for validation sets containing independent configurations sampled from the equilibrium of systems. We does not observe any outliers on these plots.

\clearpage

\section*{Green-Kubo integrals for ionic electric conductivity.}
\label{GK_IC}

We provide time dependencies of GK integration for ionic electric conductivity. Averaging over 10 trajectories (red) is employed for mean value calculation (blue). The standard deviation is reported the uncertainty. We observe that at both temperatures (900 and 1100~K) the integrals are converged. Possible, the smaller time (e.g., 10~ps) could be enough for calculations.

\begin{figure*}[h!]
	\centering
	\includegraphics[trim={0cm 0cm 0cm 0cm}, clip, width=1\linewidth]{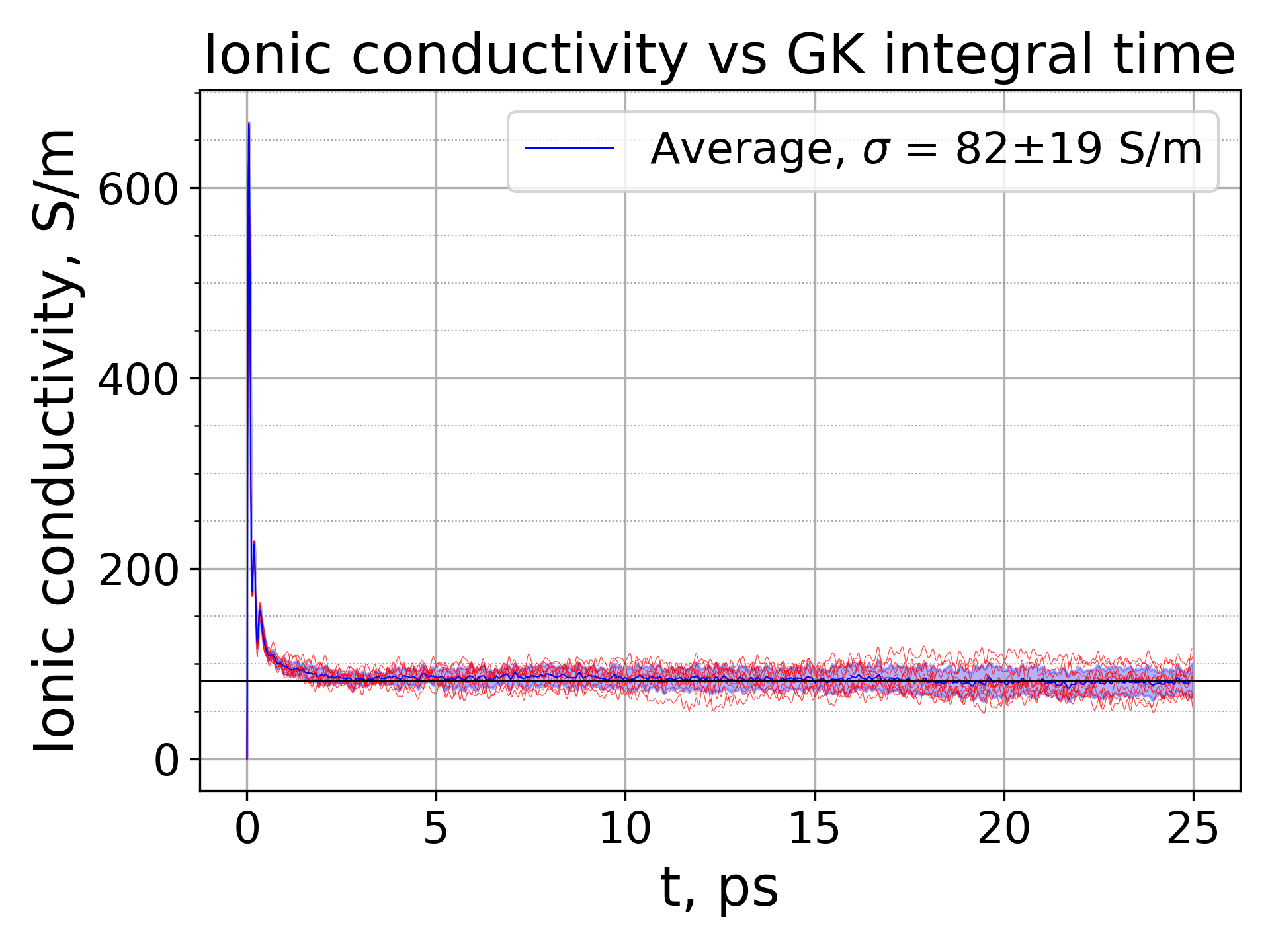}
 	\caption{Dependence of ionic conductivity on the integration time in GK method. Temperature 900 K.}
\label{fig:ACFvsT_CC_900}
\end{figure*}
\begin{figure*}[h!]
	\centering
	\includegraphics[trim={0cm 0cm 0cm 0cm}, clip, width=1\linewidth]{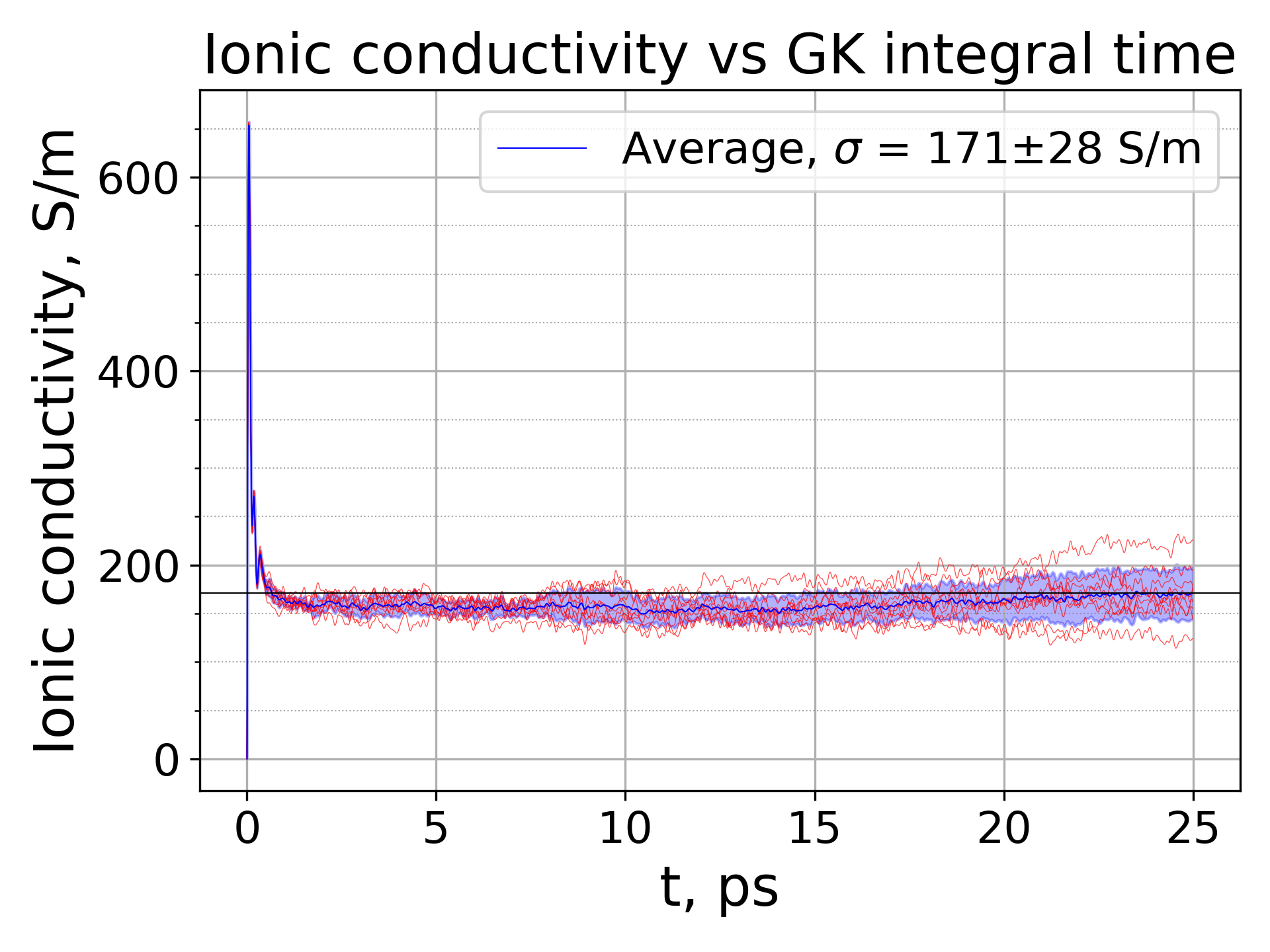}
 	\caption{Dependence of ionic conductivity on the integration time in GK method. Temperature 1100 K.}
\label{fig:ACFvsT_CC_1100}
\end{figure*}

\end{document}